\documentclass[preprint,aps,tightenlines,superscriptaddress]{revtex4}
\usepackage{graphicx}
\usepackage{bm}
\usepackage{epsfig}
\newif\ifpdf
\ifx\pdfoutput\undefined
\pdffalse 
\else
\pdfoutput=1 
\pdftrue
\fi

\newcommand{\bea}{\begin{eqnarray}}
\newcommand{\eea}{\end{eqnarray}}
\newcommand{\beq}{\begin{equation}}
\newcommand{\eeq}{\end{equation}}
\newcommand{\bay}{\begin{array}}
\newcommand{\eay}{\end{array}}
\newcommand{\vslash}{\mbox{$\not{\hspace{-1.03mm}v}$}}        

\begin{document}
\ifpdf
\DeclareGraphicsExtensions{.pdf, .jpg}
\else
\DeclareGraphicsExtensions{.eps, .jpg}
\fi

\title{
Threshold effects in excited charmed baryon decays} 

\author{Andrew E. Blechman}
\author{Adam F. Falk}
\author{Dan Pirjol}

\affiliation{Department of Physics and Astronomy, 
The Johns Hopkins University, 3400 North Charles Street, Baltimore, MD 21218}

\author{John M. Yelton}

\affiliation{Department of Physics, 
University of Florida, Gainesville, FL 32611}

\date{\today}
\begin{abstract}
\vspace{2.0cm}
\setlength\baselineskip{18pt}
Motivated by recent results on charmed baryons from CLEO and FOCUS, we reexamine
the couplings of the orbitally excited charmed baryons. 
Due to its proximity to the $\Sigma_c\pi$ threshold,
the strong decays of the $\Lambda_c^+(2593)$ are sensitive to finite width effects. 
This distorts the shape of the invariant mass spectrum in 
$\Lambda_{c1}^+\to \Lambda_c^+ \pi^+\pi^-$ from a simple Breit-Wigner resonance, 
which has implications for the experimental extraction of the $\Lambda_c^+(2593)$ 
mass and couplings. 
We perform a fit to unpublished CLEO data which gives $M(\Lambda_c^+(2593)) -
M(\Lambda_c^+) = 305.6 \pm 0.3$ MeV and $h_2^2 = 0.24^{+0.23}_{-0.11}$,
with $h_2$ the $\Lambda_{c1}\to \Sigma_c\pi$ strong coupling in the chiral Lagrangian.
We also comment on the new orbitally excited states recently observed by CLEO.
\end{abstract}
 
\maketitle

The charmed baryon system is a convenient testing ground for the ideas
and predictions of heavy quark symmetry. This is due to the rich mass spectrum 
and the relatively narrow widths of the resonances. The properties of these states
are the subject of active experimental study at both fixed target experiments
(FOCUS, SELEX, E-791) and $e^+ e^-$ machines (CLEO, BaBar, Belle). For a recent
review of the experimental situation, see Ref.~\cite{Yelton}.

In addition to the usual quantum numbers $(I, J^P)$, the charmed baryon states can be
labelled also by the spin-parity of the 
light degrees of freedom $j_\ell^{\pi_\ell}$, which are good quantum numbers
in the limit of an infinitely heavy charm quark. This property leads to
nontrivial selection rules for the strong couplings of these states to light
hadrons \cite{IsgWi}. These predictions are automatically built into an
effective Lagrangian describing the couplings of the heavy baryon states to
Goldstone bosons \cite{chPT}. 

The lowest lying charmed baryons are $L=0$ 
states and live in ${\bf \bar 3}$ and ${\bf 6}$ representations of flavor SU(3). 
It is convenient to group them together into superfields defined as in 
Ref.~\cite{chPTbar}, a vector $T_i = \frac{1+\vslash}{2}(\Xi_c^0\,, -\Xi_c^+\,,
\Lambda_c^+)_i$ for the ${\bf \bar 3}$, and a tensor $S_\mu^{ij} = \frac{1}{\sqrt3}
(\gamma_\mu + v_\mu)\gamma_5 B^{ij} + B^{*ij}_\mu$ for the ${\bf 6}$.
These superfields satisfy the constraints from heavy quark
symmetry $\vslash T = T$, $\vslash S_\mu = S_\mu$
and the  condition $\frac{1+\vslash}{2}\gamma^\mu S_\mu = 0$, which can be 
used to restrict the form of their Lagrangian interactions \cite{sfield}.
The strong couplings of the lowest lying heavy baryons are described by the effective 
Lagrangian containing two couplings $g_{1,2}$ \cite{chPTbar} (we use here the 
normalization of Ref.~\cite{PY} for these couplings)
\bea
{\cal L}_{\rm int} = \frac32 ig_1 \varepsilon_{\mu\nu\sigma\lambda}
(\bar S^\mu_{ik} v^\nu A^\sigma_{ij} S^\lambda_{jk}) - \sqrt3 g_2 \epsilon_{ijk}
(\bar T^i A_\mu^{jl} S^\mu_{kl})\,,
\eea
where $A_\mu = \frac{i}{2}(\xi^\dagger \partial_\mu \xi - \xi \partial_\mu \xi^\dagger)
= -\frac{1}{f_\pi} \partial_\mu M + \cdots$ is the usual nonlinear axial 
current of the Goldstone bosons, defined in terms of $\xi = \exp(iM/f_\pi)$ with
$f_\pi = 132$ MeV.

In this paper we focus on the negative parity $L=1$ orbitally excited 
charmed baryons. Combining the
quark spins with the $L=1$ orbital momentum gives 7 $\Lambda$-type and 7 $\Sigma$-type
states without strangeness \cite{CIK,CI} (see Table I). 
In the constituent quark model, these states fall into two distinct groups,
corresponding to the symmetric and antisymmetric irreducible representations
of $S_2$. The symmetric (antisymmetric) states are denoted in Table I with unprimed
(primed) symbols. Quark model estimates for the masses of these states \cite{CIK,CI}
suggest that symmetric states are lighter than the antisymmetric ones.
Although the permutation symmetry $S_2$ is not a true symmetry of QCD beyond the 
quark model, we will continue to refer to the higher mass charm baryon states as 
`antisymmetric', as opposed to the lower `symmetric' states.
The properties of these states were studied in the quark model in
Refs.~\cite{CIK,CI,PY,JGK} and using large $N_c$ methods in \cite{CK,largeNc,largeNc1}.

The CLEO, ARGUS and E687 Collaborations \cite{Lc1}
observed two negative parity charm baryons, $\Lambda_c^+(2593)$ and
$\Lambda_c^+(2625)$. In accordance with the expectations from the constituent
quark model, these states were
identified with the $\Lambda_{c1}(\frac12,\frac32)$ states in Table I. Their
average masses and widths are \cite{PDG}
\bea\label{lambdac1}
M(\Lambda_c^+(2593)) - M(\Lambda_c^+) &=& 308.9 \pm 0.6 \mbox{ MeV}\,,\qquad 
\Gamma(\Lambda_c^+(2593)) = 3.6^{+2.0}_{-1.3}\mbox{ MeV}\\
M(\Lambda_c^+(2625)) - M(\Lambda_c^+) &=& 341.7 \pm 0.6 \mbox{ MeV}\,,\qquad 
\Gamma(\Lambda_c^+(2625)) < 1.9\mbox{ MeV (90\% CL)}
\nonumber
\eea

\begin{center}
\begin{tabular}{|c|c|c|}
\hline
\hline
State & $(I,J)$ & $j_\ell^{\pi_\ell}$ \\
\hline
$\Lambda_{c1}(\frac12, \frac32)$ & $(0,\frac12), (0,\frac32)$ & $1^-$ \\
$\Sigma_{c0}(\frac12)$ & $(1,\frac12)$ & $0^-$ \\
$\Sigma_{c1}(\frac12, \frac32)$ & $(1,\frac12), (1,\frac32)$ & $1^-$ \\
$\Sigma_{c2}(\frac32, \frac52)$ & $(1,\frac32), (1,\frac52)$ & $2^-$ \\
\hline
$\Sigma'_{c1}(\frac12, \frac32)$ & $(1,\frac12), (1,\frac32)$ & $1^-$ \\
$\Lambda'_{c0}(\frac12)$ & $(0,\frac12)$ & $0^-$ \\
$\Lambda'_{c1}(\frac12, \frac32)$ & $(0,\frac12), (0,\frac32)$ & $1^-$ \\
$\Lambda'_{c2}(\frac32, \frac52)$ & $(0,\frac32), (0,\frac52)$ & $2^-$ \\
\hline
\hline
\end{tabular}
\begin{quote} {\bf Table I.} The quantum numbers of the expected
$p$-wave strangeless charmed baryons. The corresponding states with
strange quarks can be constructed by completing the SU(3) multiplets
to which the above states belong.
\end{quote}
\end{center}

Motivated by these data, the lowest lying states $\Lambda_{c1}(\frac12,\frac32)$
were studied in a chiral Lagrangian approach in Ref.~\cite{Cho}, where their
couplings to Goldstone bosons were first derived. These states can be grouped together
into a superfield $R_\mu^i = \frac{1}{\sqrt3}(\gamma_\mu + v_\mu) \gamma_5 R^i +
R^{(*)i}_\mu$ with $R^{(*)}_i = (\Xi_{c1}^0\,, -\Xi_{c1}^+\,, \Lambda_{c1}^+)_i$,
subject to the same constraints as the superfield $S_\mu$.

At leading order in the heavy
quark expansion, the pion couplings of these states to the sextet ground state baryons
$S_\mu$ are given by two terms, corresponding to $S-$ and $D-$wave pion emission, 
respectively
\bea
{\cal L}_{\rm int} = h_2 \epsilon_{ijk} \bar S_\mu^{kl} v_\nu A^\nu_{lj} R_\mu^i +
ih_8 \epsilon_{ijk} \bar S_\mu^{kl} \left( {\cal D}^\mu A^\nu + 
{\cal D}^\nu A^\mu + \frac23 
g^{\mu\nu} (v\cdot {\cal D})(v\cdot A)\right)_{lj} R_\nu^i + \mbox{h.c.}
\eea
with the covariant derivative ${\cal D}_\mu A_\nu = \partial_\mu A_\nu + [V_\mu\,,
A_\nu ]$ and $V_\mu = \frac12 (\xi^\dagger \partial_\mu \xi + \xi \partial_\mu
\xi^\dagger )$.
This formalism was extended to the other $p-$wave charmed baryons in Table I in
Refs.~\cite{PY,CF}, where prospects were given for their discovery. A total of 
6 $S-$wave and 8 $D$-wave couplings are required for a complete description of the
strong couplings of the states in Table I. 

Knowledge of the pion couplings $h_2, h_8$ of the lowest orbital excitations 
$\Lambda_{c1}(\frac12,\frac32)$ will provide information about the other excited
baryons, and could thus help guide the search for the missing states. For example,
assuming SU(3) symmetry, the widths of the orbitally excited charm
baryons containing strange quarks $\Xi'_{c1}(\frac12,\frac32)$ can be predicted
\cite{CF,PY}, with results in good agreement with the
CLEO data on $\Xi'_{c1}(\frac12)$ \cite{CLEOXic1/2} and 
$\Xi'_{c1}(\frac32)$ \cite{CLEOXic3/2}.
Furthermore, in the constituent quark model, the couplings of all unprimed states
in Table I can be shown to be related to $h_2, h_8$ \cite{PY,JGK}. Assuming that 
the masses of these states are known, these relations can be therefore used to 
predict the decay modes and widths of all these states. Finally, once determined
in the charm system, the same couplings would also give the properties of the
excited bottom baryons.
Clearly, a precise determination of the two couplings $h_2, h_8$ is of great interest.

There are a few issues which complicate such a determination, following
from the peculiarities of the actual mass spectrum. The states
$\Lambda_{c1}(\frac12,\frac32)$
are observed through their 3-body decays in the $\Lambda_c^+ \pi^+\pi^-$ channel. 
These are resonant decays, proceeding through intermediate $\Sigma_c^{(*)}\pi$ states.
The masses, and recently the widths of the $\Sigma_c$ baryons have been measured 
by the FOCUS \cite{FOCUS_sigma} and CLEO \cite{CLEO_sigma} Collaborations. 
The average results of these measurements are \cite{PDG}
\bea\label{sigmac}
& &M(\Sigma_c^{++})-M(\Lambda_c^+) = 167.67\pm 0.15\mbox{ MeV}\,,\qquad
\Gamma(\Sigma_c^{++}) = (2.05^{+0.41}_{-0.38}\pm 0.38) \mbox{ MeV}\quad  \\
& &M(\Sigma_c^{+})-M(\Lambda_c^+) = 166.4\pm 0.4\mbox{ MeV}\,,\qquad\qquad
\Gamma(\Sigma_c^{+}) \leq 4.6 \mbox{ MeV} (90\%\mbox{ CL})\nonumber\\
& &M(\Sigma_c^{0})-M(\Lambda_c^+) = 167.32\pm 0.15\mbox{ MeV}\,,\qquad\quad
\Gamma(\Sigma_c^{0}) = (1.55^{+0.41}_{-0.37}\pm 0.38) \mbox{ MeV}\nonumber
\eea
In the heavy quark limit, the only allowed resonant channels are $\Lambda_{c1}(\frac12)
\to [\Sigma_c \pi]_S , [\Sigma^*_c \pi]_D$, and $\Lambda_{c1}(\frac32)\to
[\Sigma_c \pi]_D , [\Sigma^*_c \pi]_{S,D}$, where the subscript denotes the orbital 
angular momentum. 
From (\ref{lambdac1}) and (\ref{sigmac}) it follows that the dominant $S-$wave 
decays of the $\Lambda_{c1}(2593)$ proceed very close to threshold. Furthermore,
the available energy in the decay is comparable or less than the width of the 
decaying state
\bea
\Lambda_{c1}(2593) - \left[
\begin{array}{c}
(\Sigma_c^0(2455) + \pi^+ ) \\ 
(\Sigma_c^{++}(2455) + \pi^- )\\
\end{array}
\right]
\sim
\left(
\begin{array}{c}
2 \mbox{ MeV} \\
1.7 \mbox{ MeV} \\
\end{array}
\right)
\leq \Gamma(\Lambda_{c1}^+(2593))
\eea
On the other hand, the decay into the $\Sigma^+\pi^0$ channel takes place $\sim 7.5$ 
MeV above threshold, such that it turns out to dominate the width of the 
$\Lambda_{c1}(2593)$.

The situation with the spin-$\frac32$ state $\Lambda_c(2625)$ is somewhat different.
For this case, the decay is dominated by the $D-$wave channel $[\Sigma_c\pi]_D$,
which is well above threshold \mbox{$(\sim 45$ MeV)}, while the $S-$wave accessible modes 
$[\Sigma_c^*\pi]_S$ lie about $30$ MeV below threshold and are thus nonresonant. 

This suggests that finite width effects are important in the $\Lambda_c(2593)$
decays. The situation is somewhat
similar to $e^+ e^- \to t\bar t$ production close to threshold, which is mediated by
a very broad toponium resonance. The net effect is a distortion of the shape of the
invariant mass spectrum in $\Lambda_{c1}(2593)\to \Lambda_c^+\pi^+\pi^-$ from a simple
Breit-Wigner shape. The resulting line shape depends both on the unknown couplings 
$h_{2,8}$ and on the masses and widths of the intermediate $\Sigma_c$ states.
This should be taken into account for the extraction of the
mass and width of the $\Lambda_{c1}(2593)$. The purpose of this paper is to present
a detailed calculation of these effects.

Consider the amplitude for producing the $\Lambda_{c1}$ resonance, followed by
its decay to a 3-body state $\Lambda_{c1}^+\to \Lambda_c^+\pi\pi$, of total momentum
$p_\mu = M_{\Lambda_c^+} v_\mu + k_\mu$ and 
invariant mass $M(\Lambda_c^+\pi\pi) = \sqrt{p^2(\Lambda_c\pi\pi)}$. 
This is written in the factorized form
\bea\label{ampl}
A(i\to \Lambda_{c1}X \to \Lambda_c^+\pi\pi X) = 
\frac{i}{\Delta - \Delta_{\Lambda_{c1}} +
i\Gamma_{\Lambda_{c1}}(\Delta)/2} [\bar U(\Delta)\frac{1+\vslash}{2} V(\Delta, X)]\,,
\eea
where $\Delta  = v\cdot k = M(\Lambda_c^+\pi\pi) - M(\Lambda_c^+)$ 
is the residual energy of the propagating resonance $\Lambda_c(2593)$ and
$\Delta_{\Lambda_{c1}} = M(\Lambda_{c1}) - M(\Lambda_{c}^+)$.
$U_\alpha(\Delta)$ and $V_\alpha(\Delta,X)$ are spinor amplitudes parameterizing the
decay $\Lambda_{c1}^+\to \Lambda_c^+\pi\pi$ and its production,
respectively. $U_\alpha(\Delta)$ depends on the momenta and spins of the $\Lambda_c
\pi\pi$ state, and is calculable in heavy hadron chiral perturbation theory for
values of the residual energy $\Delta \ll 1$ GeV. On the other hand, not much is
known about the production spinor $V_\alpha(\Delta,X)$, which depends on all the
details of the production process.

Squaring the amplitude (\ref{ampl}), adding the phase space factors and summing
over the unobserved states $X$, one finds
the following expression for the $\Lambda_c^+\pi\pi$ production cross-section 
as a function of the invariant mass $\Delta$:
\bea\label{xsec}
\frac{d\sigma(\Delta)}{d\Delta} \sim
\frac{1}{(\Delta-\Delta_{\Lambda_{c1}})^2 + \Gamma_{\Lambda_{c1}}(\Delta)^2/4}
\left[ \bar U(\Delta) \frac{1+\vslash}{2}\omega(\Delta) \frac{1+\vslash}{2}
U(\Delta)\right] \mbox{dLips}(\Lambda_{c1} \to \Lambda_c^+\pi\pi)
\eea
We have introduced here the density matrix $\omega_{\alpha\beta}(\Delta)$ 
parameterizing
the production of a $\Lambda_{c1}$ resonance in the process $i \to \Lambda_{c1} X$
\bea
\omega_{\alpha\beta}(\Delta) \equiv
\sum_X \int d\mu(X) V_\alpha(\Delta,X) \bar V_\beta(\Delta,X) 
(2\pi)^4 \delta(p_i - p_X - p_{\Lambda_{c1}}) 
\eea
The matrix $\omega$ depends on the resonance momentum $p_{\Lambda_{c1}}$ and 
details of the experimental setup such as the total beam momentum and polarization. 
Fortunately, the spin structure of the matrix $\omega$ is not
required if one sums over the spins and momenta of the final decay products
in $\Lambda_{c1}^+ \to \Lambda_c^+ \pi\pi$. If this is done, the amplitudes in 
Eq.~(\ref{xsec}) can be written as
\bea
\sum_{s_{\Lambda_c}}\int \mbox{dLips}(\Lambda_{c1} \to \Lambda_c^+\pi^+\pi^-)
U_\alpha(\Delta) \bar U_\beta(\Delta) = \left(\frac{1+\vslash}{2}\right)_{\alpha\beta}
\Gamma(\Lambda_{c1}^+ \to \Lambda_c^+\pi^+\pi^-)
\eea
Inserting this into (\ref{xsec}) one finds that the production
cross section as a function of invariant mass takes the factorized form
\bea\label{10}
\frac{d\sigma(\Delta)}{d\Delta} \sim
\mbox{Tr }\left[\frac{1+\vslash}{2}\omega(\Delta)\right]
\frac{\Gamma(\Lambda_{c1}^+ \to \Lambda_c^+\pi^+\pi^-)}
{(\Delta-\Delta_{\Lambda_{c1}})^2 + \Gamma^2_{\Lambda_{c1}}(\Delta)/4}
\eea
The dependence on $\Delta$ introduced by the production factor $\mbox{Tr }[
\frac{1+\vslash}{2}\omega(\Delta)]$ is unknown, and it can be expected to
introduce a slow variation with a characteristic scale $\sim \Lambda_{QCD}$.
This can be neglected when compared with the much faster variation of the denominator. 
The width $\Gamma(\Delta)$ in the numerator is equal to the spin-averaged 
partial width of a
$\Lambda_{c1}$ resonance of mass $\Delta + M(\Lambda_c^+)$ into a specific channel,
e.g. $\Lambda_c^+\pi^+\pi^-$, while the width in the denominator 
$\Gamma_{\Lambda_{c1}}(\Delta)$ sums over all allowed channels. 
These decay widths are given explicitly by \cite{PY} 
\bea
\Gamma_{+-} \equiv
\Gamma(\Lambda_{c1}^+ \to \Lambda_c^+\pi^+\pi^-) &=& \frac{g_2^2}{16\pi^3 f_\pi^4}
M_{\Lambda_{c}^+} \int dE_1 dE_2 \left\{ \vec p_2\,^2 |A(E_1,E_2)|^2 + \vec p_1\,^2 
|B(E_1,E_2)|^2\right.\nonumber\\
& &\label{+-}
\left. 
+ 2\vec p_1 \cdot \vec p_2 \mbox{Re }[A(E_1,E_2) B^*(E_1,E_2)]\right\}
\eea
where $E_1, E_2$ are the pion energies in the rest frame of the $\Lambda_{c1}$ resonance and
we have defined
\bea
& &A(E_1, E_2) = \frac{h_2 E_1}{\Delta - \Delta_{\Sigma_c^0} - E_1 + i\Gamma_{\Sigma_c^0}/2}\\
& &\qquad + h_8
\left( - \frac{\frac23 \vec p_1\,^2}{\Delta - \Delta_{\Sigma_c^{*0}} - E_1 + i\Gamma_{\Sigma_c^{*0}}/2}
+ \frac{2\vec p_1\cdot \vec p_2}{\Delta - \Delta_{\Sigma_c^{*++}} - E_2 + i\Gamma_{\Sigma_c^{*++}}/2}
\right)\nonumber\\
& &B(E_1, E_2; \Delta_{\Sigma_c^{(*)0}}, \Delta_{\Sigma_c^{(*)++}}) =
A(E_2, E_1; \Delta_{\Sigma_c^{(*)++}}, \Delta_{\Sigma_c^{(*)0}})
\eea
The decay rate $\Gamma(\Lambda_{c1}^+\to \Lambda_c^+\pi^0\pi^0)$ is given by a similar 
relation, with an additional factor of $1/2$ to account for the identical pions 
in the final state, and with the replacements 
$\Delta_{\Sigma_c^{(*)++}}, \Delta_{\Sigma_c^{(*)0}} \to \Delta_{\Sigma_c^{(*)+}}$.

In these expressions we work at leading order in the $1/m_c$ expansion in matrix 
elements, but use the exact 3-body phase space. This procedure includes formally
subleading contributions in the $1/m_c$ expansion, which are however enhanced by 
kinematics and are required for reproducing the data in other similar situations 
\cite{MLW}.
We neglect the radiative decay channel $\Lambda_{c1}^+\to \Lambda_c^+\gamma$, 
which is expected to contribute about 20 keV to the total width \cite{CK}.

\begin{figure}[t!]
\begin{center}
  \includegraphics[width=3.2in]{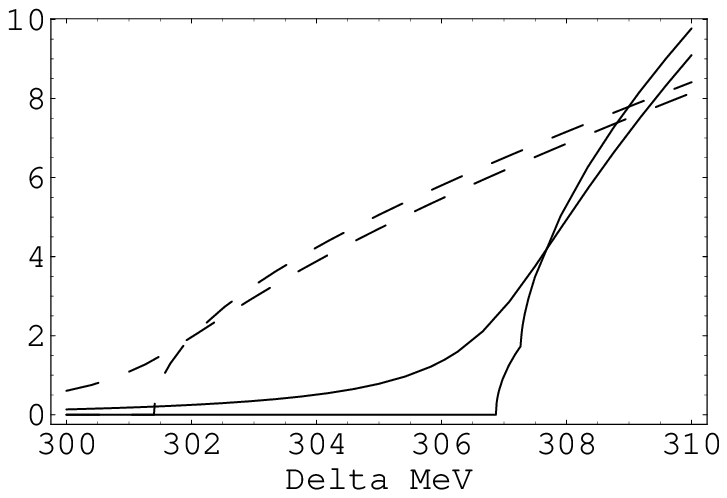}
  \includegraphics[width=3.2in]{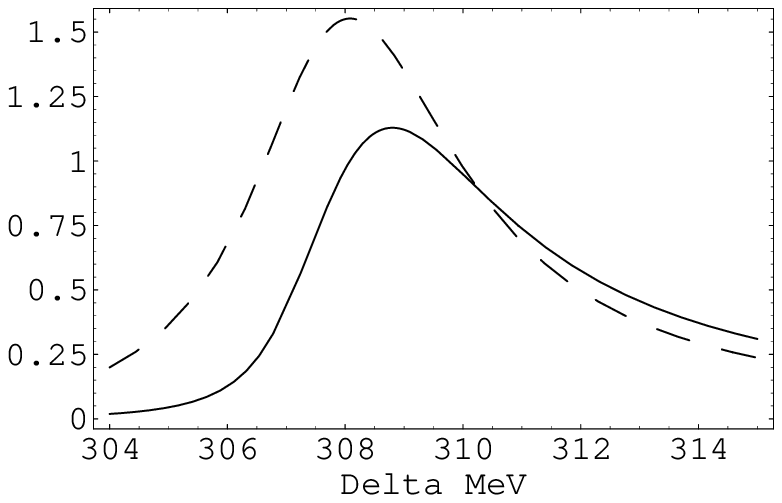}\\
\hspace{0.5cm} (a) \hspace{7cm} (b)
\end{center}
{\caption{
(a) The partial mass-dependent width of the $\Lambda_c(2593)$ in the
$\Lambda_c^+\pi^+\pi^-$ channel ($g_2^2 a_{+-}(\Delta)$ - solid line) and 
in the $\Lambda_c^+\pi^0\pi^0$ channel ($g_2^2 a_{00}(\Delta)$ - dashed line), 
as a function of 
$\Delta = M(\Lambda_c^+\pi\pi)-M(\Lambda_c^+)$, with $g_2^2=0.34$;
the curves with sharp thresholds are computed in the narrow width approximation
(Eqs.~(\ref{16}), (\ref{17})) and are independent on $g_2$;
(b) The $\Lambda_c^+(2593)$ resonance shape as seen in the
$\Lambda_c^+\pi^+\pi^-$ channel (solid curve) and
in the $\Lambda_c^+\pi^0\pi^0$ channel (dashed curve).
The results in (b) correspond to 
$\Delta_{\Lambda_{c1}}= 309$ MeV and $h_2^2 = 0.3$.}} 
\end{figure}

After integration over the Dalitz plot, the decay width (\ref{+-}) can be written as
\bea\label{g2a}
\Gamma_{+-}(\Delta) = g_2^2\left\{
h_2^2 a_{+-}(\Delta) + h_8^2 b_{+-}(\Delta) + 2h_2 h_8 c_{+-}(\Delta)\right\}\,.
\eea
A similar result is obtained for the rate into $\Lambda_c^+\pi^0\pi^0$ with 
coefficients $a_{00}, b_{00}, c_{00}$. The coupling $g_2$ appears here both 
explicitly, and
implicitly through the $\Sigma_c^{(*)}$ widths in the denominators of $A(E_1, E_2)$ 
and $B(E_1, E_2)$. These are given by
\bea
\Gamma(\Sigma_c^{(*)}) = \frac{g_2^2}{2\pi f_\pi^2}
\frac{M_{\Lambda_c}}{M_{\Sigma_c^{(*)}}} |\vec p_\pi\,|^3\,.
\eea
Using the observed masses this gives $\Gamma(\Sigma_c^{++,+,0}) = \{6.15\,,
7.06\,, 6.01\} g_2^2 $ MeV, and $\Gamma(\Sigma_c^{++*,+*,0*}) = \{47.9\,,
47.4\,, 46.3\} g_2^2 $ MeV. The extracted values for $g_2$ from the 
$\Sigma_c$ and $\Sigma_c^*$ experimental widths are somewhat different:
$\langle g_2^2\rangle_{\Sigma_c} = 0.25 \pm 0.17$, and 
$\langle g_2^2\rangle_{\Sigma_c^*} = 0.33 \pm 0.15$, which can be
attributed to an $1/m_c$ effect. Although the uncertainty in this coupling is rather
large, $g_2^2 = 0.29\pm 0.23$, the resulting effect on our predictions (\ref{g2a})
is very small, because they are very close to the narrow-width case for the
$\Sigma_c$ (see the discussion around Eqs.~(\ref{16}), (\ref{17})).

\begin{figure}[t!]
\begin{center}
\includegraphics[width=5in]{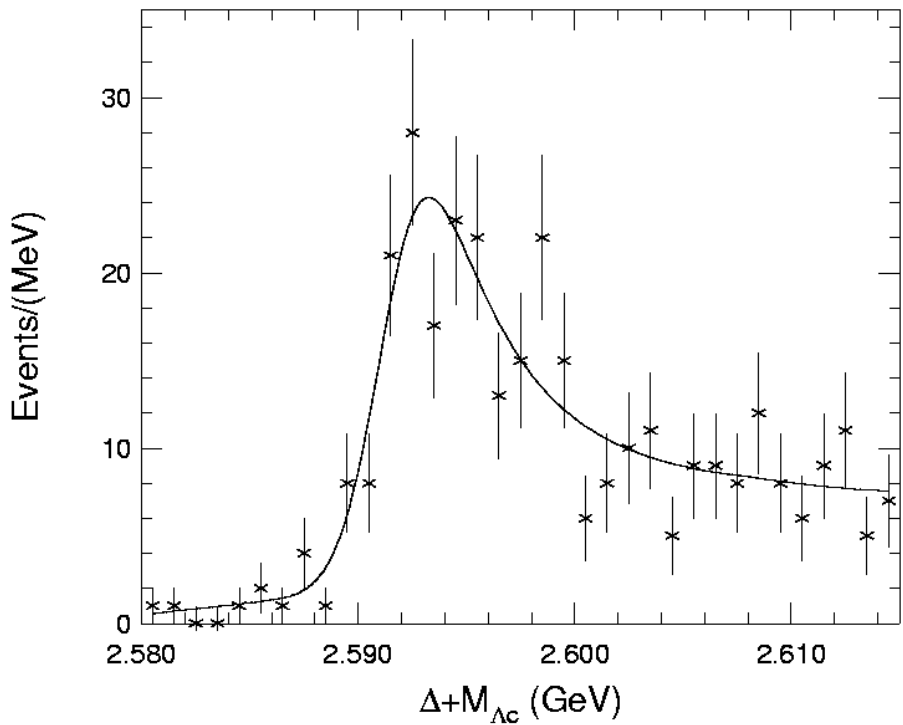}
\end{center}
{\caption{
Fit to the invariant mass spectrum in $\Lambda_c^+(2593)\to \Lambda_c^+\pi^+\pi^-$
as explained in the text.}}
\end{figure}

Our main interest here is in the functional dependence of $a_{+-,00}(\Delta)$, which 
dominate numerically the rates $\Gamma_{+-,00}$. These coefficients are plotted
in Fig.~1(a) as functions of $\Delta$; the qualitative features of these 
curves can be understood without a detailed computation, as follows.
The coefficients $a(\Delta)$ give the partial widths into the $[\Sigma_c\pi]_S$ channel,
which start at threshold $\Delta = 2M(\pi^+)$, and rise slowly up to the threshold for
production of $[\Sigma_c^{0}\pi^{+}]_S$ and $[\Sigma_c^{++}\pi^{-}]_S$ at 
$\Delta = 306.9$ MeV and $\Delta = 307.2$ MeV,
respectively. Above this threshold, the rate rises much faster, which explains the 
`kink' seen in Fig.~1(a) in the $\pi^+\pi^-$ channel. On the other hand, the threshold 
in the neutral pion channel lies lower, at $\Delta=301.4$ MeV, corresponding to the 
opening of the $[\Sigma_c^+\pi^0]_S$ channel. Since the central value of the 
$\Lambda_{c1}$ mass lies around $ 307$ MeV, the rapid variation of $a_{+-}(\Delta)$ 
in this region will likely affect the extraction of $\Delta_{\Lambda_{c1}}$.

It is instructive to compare these results with those obtained in the 
narrow width approximation, where the mass-dependent partial widths in 
(\ref{+-}) are approximated with 2-body widths \cite{CF}
\bea\label{16}
\Gamma_{\rm NW}(\Lambda_{c1}^+ \to \Lambda_c^+\pi^+\pi^-) &=& 
\Gamma(\Lambda_{c1}^+ \to \Sigma_c^0\pi^+) + 
\Gamma(\Lambda_{c1}^+ \to \Sigma_c^{++}\pi^-) = a(\pi^\pm) |\vec p_\pi\,|\\
\label{17}
\Gamma_{\rm NW}(\Lambda_{c1}^+ \to \Lambda_c^+\pi^0\pi^0) &=& 
\Gamma(\Lambda_{c1}^+ \to \Sigma_c^+\pi^0)  = a(\pi^0) |\vec p_\pi\,|
\eea
where $\vec p_\pi$ is the pion momentum in  $\Lambda_{c1}\to \Sigma_c\pi$ decays.
Neglecting isospin violation in the $\Sigma_c$ masses,
the $a(\pi)$ parameters are given in the heavy quark limit by
\bea
a(\pi^\pm) = \frac{h_2^2}{\pi f_\pi^2} \frac{M_{\Sigma_c}}{M_{\Lambda_{c1}}}
E_\pi^2\,,\qquad a(\pi^0) = \frac12 a(\pi^\pm)\,.
\eea
In the limit $g_2\to 0$, the exact result (\ref{+-}) reduces to the narrow width
approximation in Eqs.~(\ref{16}) and (\ref{17}), that is $\Gamma \to \Gamma_{\rm NW}$.
As one can see from Fig.~1(a), the narrow width results give a good approximation 
to the exact widths (computed with $g_2^2=0.34$), for $\Delta$ not too close to 
threshold.

In Fig.~1(b) we show invariant mass distributions
$\Delta = M(\Lambda_c^+ \pi\pi) - M(\Lambda_c^+)$ in $\Lambda_c^+(2593)$ decays,
in both charged and neutral pions channels. 
The shape of the invariant mass distribution in the charged pions channel 
$\Lambda_c^+\pi^+\pi^-$ is distorted towards larger values of $\Delta$ compared to a
simple Breit-Wigner curve.
In particular, extractions of the $\Lambda_c^+(2593)$ parameters from the charged
pions channel alone could overestimate the mass of this resonance by a few MeV,
which is larger than the present $1\sigma$ uncertainty (\ref{lambdac1}) on this 
parameter.
These effects are not present in the neutral pions channel, for which the 
shape of the mass spectrum comes closer to a pure Breit-Wigner resonance. 

The first observation of the $\Lambda_c^+\pi^0\pi^0$ mode has been presented in
unpublished CLEO data \cite{thesis}, where the
corresponding invariant mass distribution was used to extract the mass of the
$\Lambda_c^+(2593)$. The result is lower than that obtained from the
$\Lambda_c^+\pi^+\pi^-$ channel (\ref{lambdac1}), in agreement with our expectations,
\bea\label{pi0pi0}
[M(\Lambda_c^+(2593)) - M(\Lambda_c^+)]_{\Lambda_c\pi^0\pi^0} &=& 
306.3 \pm 0.7 \mbox{ MeV}\,.
\eea
Experimental difficulties connected with the low $\pi^0$ detection efficiency 
could limit the precision of such a determination.
We propose therefore that the shape of the $\Lambda_c^+ \pi^+\pi^-$ invariant mass 
spectrum be fit to the distribution (\ref{10}) with parameters 
$(\Delta_{\Lambda_{c1}},h_2)$ (instead of a Breit-Wigner curve with 
parameters $(\Delta_{\Lambda_{c1}},\Gamma)$).

In Fig.~2 we show the results of such a fit, performed using the CLEO data 
presented in \cite{thesis} (see Fig.~5.5 in this reference), including detector
resolution effects.
The parameters of the $\Lambda_c(2593)$ resonance extracted from this fit 
are \footnote{The data shown in Fig.~2 was obtained in Ref.~\cite{thesis} by adding the
measured mass difference to a fixed $\Lambda_c^+$ mass of 2286.7 MeV. Thus, for
consistency, we subtracted this value from our fitted mass to obtain the result
(\ref{final}).}
\bea\label{final}
M(\Lambda_c^+(2593)) - M(\Lambda_c^+) = 305.6\pm 0.3 \mbox{ MeV}\,,\qquad
h_2^2 = 0.24^{+0.23}_{-0.11}\,,
\eea
and correspond to a resonance mass in reasonably good agreement with (\ref{pi0pi0}).
A conventional fit of this same data using a Breit-Wigner function, yields a
mass difference of around 308 MeV, in agreement with the published measurements
\cite{Lc1}.
Note that the threshold effects effectively lower the resonance mass  
(\ref{final}) compared with the previous determinations (\ref{lambdac1}). Our treatment
also leads to a reduction in the uncertainties connected with the poorly measured
$\Sigma_c$ widths. The result for the coupling $h_2^2$ is somewhat lower than
previous determinations of this coupling \cite{CF} ($h_2^2 = 0.30^{+0.21}_{-0.14}$)
and \cite{PY} ($h_2^2 = 0.33^{+0.20}_{-0.13}$).


Finally, we comment on the recent evidence by the CLEO Collaboration
\cite{CLEO_new} for new charmed baryon states, lying above the $\Lambda_c(2593)$
and $\Lambda_c(2625)$. The lower signal $X_1$ is relatively broad, while the 
higher peak $X_2$ is narrow, with masses and widths
\bea\label{lambdac0}
M(X_1) - M(\Lambda_c^+) &=& 480.1 \pm 2.4  \mbox{ MeV}\,,\qquad 
\Gamma(X_1) = 20.9\pm 2.6\mbox{ MeV}\\
M(X_2) - M(\Lambda_c^+) &=& 595.8 \pm 0.8 \mbox{ MeV}\,,\qquad 
\Gamma(X_2) = 4.2 \pm 0.7\mbox{ MeV}
\nonumber
\eea

The higher narrow peak $X_2$ has been identified in \cite{CLEO_new} with 
the antisymmetric state $\Lambda'_{c0}(\frac12)$, while the
lower broad peak $X_1$ has been interpreted as the overlap of the peaks
corresponding to the two
$\Sigma_{c2}(\frac32,\frac52)$ states in Table I, which can both decay to 
$[\Lambda_c^+\pi]_D$ and $[\Sigma_c^{(*)}\pi]_D$.
 In the heavy quark limit,
the $\Lambda'_{c0}(\frac12)$ does not decay to any of the lower lying charmed baryons.
The only such mode allowed by isospin and heavy quark symmetry is to $\Xi_c K$,
which is however kinematically forbidden. According to the interpretation of
\cite{CLEO_new}, it can be seen in the $\Sigma_c\pi$ channel through its mixing with
the $\Lambda'_{c1}(\frac12)$ at subleading order in $1/m_c$. This is consistent with 
the experimental observation of $\Sigma_c\pi$ resonant substructure (which is accessible
in $\Lambda'_{c1}(\frac12)\to [\Sigma_c\pi]_S$), but not of $\Sigma_c^*\pi$, which 
can only proceed through $D-$wave $\Lambda'_{c1}(\frac12)\to
[\Sigma_c^*\pi]_D$ and is therefore expected to be suppressed.

The mass measurement (\ref{lambdac0}) of the $X_2$ state shows that it lies above the
threshold for $[ND]$ ($\Delta=524$ MeV for $nD^+$ and $\Delta = 518$ MeV for
$pD^0$). Both the $\Lambda'_{c0}(\frac12)$ and $\Lambda'_{c1}(\frac12)$
can decay to this mode in an orbital
$S-$wave in the heavy quark limit (which would give thus the dominant decay
mode in the absence of $1/m_c$ effects). Therefore we would like to suggest that 
one search for the $X_2$ state also in the $ND$ channel, where it should show up 
as well. Observing such a signal would definitely rule out alternative
interpretations of this state as $\Sigma_{c2}(\frac32,\frac52)$ or $\Lambda'_{c1}
(\frac32), \Lambda'_{c2}(\frac32,\frac52)$, which can decay to $ND$ only in $D-$wave.

In conclusion, we have discussed in this paper the impact of threshold effects on the 
determination of the $\Lambda_c^+(2593)$ parameters from its strong decays into 
$\Lambda_c^+ \pi\pi$, and we have presented theory motivated fits of the mass and 
couplings of this state. Our results suggest that the excitation energy of the 
$\Lambda_c^+(2593)$ is about 2-3 MeV lower than obtained in previous determinations.

\hspace{0.5cm}

This research was supported by the U.S. NSF Grant PHY-9970781 (A.E.B., A.F.F. 
and D.P.) and by the DOE grant DE-FG02-97ER41029 (J.M.Y.).

\end{document}